\documentclass[sigconf,nonacm]{acmart}

\renewcommand\footnotetextcopyrightpermission[1]{}
\usepackage{booktabs}
\usepackage{subcaption}
\usepackage{graphicx}
\usepackage{xcolor}
\usepackage{amsmath}
\usepackage{multirow}
\usepackage{setspace}
\usepackage{lineno}
\usepackage{hyperref}

\newif\ifdraft
\draftfalse

\newcommand{\nb}[2]{
	{
		{\color{black}{
				\fbox{\bfseries\sffamily\scriptsize#1}
				{\sffamily$\triangleright~${\it\sffamily #2}$~\triangleleft$}
	}}}
}

\ifdraft
\newcommand\vahid[1]{\nb{Vahid}{\color{blue}#1}}
\newcommand\grex[1]{\nb{Gregorio}{\color{red}#1}}

\newcommand\reviewer[1]{\nb{Reviewer}{\color{red}#1}}
\newcommand{\fixme}[1]{{\textcolor{red}{[FIXME] #1}}\xspace}

\else
\usepackage[disable]{todonotes}
\newcommand\vahid[1]{}
\newcommand\grex[1]{}
\newcommand\reviewer[1]{}
\newcommand{\fixme}[1]{}

\fi

\usepackage[inline]{enumitem}

\title{SmartDoc: A Context-Aware Agentic Method Comment Generation Plugin}

\author{Vahid Etemadi}
\affiliation{%
	\institution{Independent Researcher}
	\city{Shiraz}
	\country{Iran}
}
\email{vetemadi87@gmail.com}

\author{Gregorio Robles}
\affiliation{%
	\institution{Universidad Rey Juan Carlos}
	\city{Madrid}
	\country{Spain}
}
\email{grex@gsyc.urjc.es}

\begin{document}

\begin{abstract}
\textbf{Context}: The software maintenance phase involves many activities such as code refactoring, bug fixing, code review or testing.
Program comprehension is key to all these activities, as it demands developers to grasp the knowledge (e.g., implementation details) required to modify the codebase.
Methods as main building blocks in a program can offer developers this knowledge source for code comprehension.
However, reading entire method statements can be challenging, which necessitates precise and up-to-date comments.
\textbf{Objective}: We propose a solution as an IntelliJ IDEA plugin, named SmartDoc, that assists developers in generating context-aware method comments.
\textbf{Method}: This plugin acts as an Artificial Intelligence (AI) agent that has its own memory and is augmented by target methods' context.
When a request is initiated by the end-user, the method content and all its nested method calls are used in the comment generation.
At the beginning, these nested methods are visited and a call graph is generated. This graph is then traversed using depth-first search (DFS), enabling the provision of full-context to enrich Large Language Model (LLM) prompts.
\textbf{Result}: The product is a software, as a plugin, developed for Java codebase and installable on IntelliJ IDEA.
This plugin can serve concurrently for methods whose comments are being updated , and it shares memory across all flows to avoid redundant calls.
o measure the accuracy of this solution, a dedicated test case is run to record SmartDoc generated comments and their corresponding ground truth.
For each collected result-set, three metrics are computed, BERTScore, BLEU and ROUGE-1.
These metrics will determine how accurate the generated comments are in comparison to the ground truth.
\textbf{Result}: The obtained accuracy, in terms of the precision, recall and F1, is promising, and lies in the range of \textbf{0.80} to \textbf{0.90} for BERTScore.
In addition, a feedback mechanism is developed to enable receiving users' opinion as satisfaction grade and their text.
\end{abstract}


\maketitle

\section{Introduction}
\label{sec:intro}
Code comprehension is a prerequisite for making software changes, as it influences the developer's perception of the code to be changed.
This perception is necessary for developers to make their changes in a way to avoid breaking the functionality (especially in a low test-coverage rate).
These changes could be refactoring the code, adding new features, or fixing a bug.
Occasionally, code reviewers must understand and read the code before they can offer their advice, accept or reject a pull request.
The level of understanding of the code to be changed also affects developer productivity, reducing the likelihood of missed deadlines.

There is a process behind understanding the code, in which code summarization and code comments can play a key role.
For example, at the method level, the comment provided would help software developers to facilitate this understanding process, especially if there are nested calls in the body, or it is unapproachably long~\cite{huang2023comparative}.
Given this complexity, these methods are usually left commentless or sometimes become outdated when a change is applied to the method but not to its comment~\cite{huang2025your}.
Over time, this process can become detrimental to the project if developers who wrote that code are no longer in the team~\cite{etemadi2022task}.
In all these cases, a developer who touches that block of code (i.e., the method) for the first time will often need more time to make the necessary changes.
This is because incomplete understanding of the code may lead to introducing code smells or transitive technical debt~\cite{niazi2023investigating}.

In this paper, we present SmartDoc, an IntelliJ IDEA plugin designed to help developers who need to quickly grasp a method's functionality from its comment.
This tool consumes a remote/local LLM server by providing appropriate context (Retrieval Augmented Generation (RAG)), and offers an AI-generated comment for the given method.
This process is done by an implemented agent that handles each user request.
After initiating the request as an action, the handler which is linked to agent core receives the request carrying target method metadata.
It then goes through several key processes to deliver eventual generated structurally valid comment.
Internally, the agent traverses nested method calls to enrich its prompt that is finally submitted to the LLM service.
In this way, with minimal effort and by harnessing the power of LLMs, a developer can save time and effort devoted to making the intended changes.

%
%
%

The rest of the paper is organized as follows: in Section~\ref{sec:rel-works}, related works are investigated.
In Section~\ref{sec:smartdoc}, the conceptual design of SmartDoc is presented, and the architectural details of the created tool are delivered.
In Section~\ref{sec:res+eval}, we provide the evaluations applied to the generated comments by SmartDoc, and Section~\ref{sec:dis+conc} discusses the tool and concludes this paper.

\section{Related works}
\label{sec:rel-works}



Program comprehension refers to the ability to understand code at any level of granularity --whether for maintenance, debugging, review, or development-- thereby enabling effective reasoning and software modification~\cite{maalej2014comprehension}.
Understanding the code may become challenging as developers have to spend a lot of time if they do not have access to the relevant resources.
The situation can become even more complex when developers work with a tightly coupled codebase.
According to the Developer Ecosystem Survey conducted by JetBrains, 42\% of developers spend less than 1 hour each day on manual code reviews.
Other 45\% spend 1-2 hours per day.
Much of this effort likely is devoted to understanding and comprehending the code.
What if this time could be saved through the automatic generation of method comments designed to convey the functionality of each method?

Generative AI, and specifically LLMs are invented to help developers to become more productive and precise during software development~\cite{sauvola2024future}.
Code summarisation is the task of interpreting a given block of code into an understandable and easy to follow text, usually more human readable text developed close to natural language format~\cite{ahmad2020transformer}.
It serves various purposes, including as an approach for describing a method’s functionality.
While code comments are not always a natural interpretation of the code, they can convey the method intent with less effort.
One of the key features of recently appeared LLMs is their fundamental design feature that enable an effective summaraization based on the user prompt~\cite{sun2024source}.
These types of transformer-based models are capable of generating desired output given an engineered prompt.
LLMs can be instructed to produce desired output, however, they are featured as being non-deterministic.
Non-deterministic behavior means that in each try with the same input directed to a specific single model, different and still correct responses are expected.
In addition, when it is related to producing documents, responses should follow specific structure, opening a new topic about structured output~\cite{liu2024we}.

LLMs are resource intensive when trained in terms of the required infrastructure.
However, they can be fine-tuned for specific use case, e.g, a dedicated codebase~\cite{lin2024data}.
But a fine-tuned model that is expected to deal with many types of prompts needs a high volume of resources, making the deployment more costly.
In this case, some techniques can be effective, such as:
\begin{itemize}
	\item Using a caching system, a local, structurally-organized memory that avoids re-submitting prompts and keeps former responses for possible hits.
	\item Include limitations for number of tokens submitted for a given prompt or an engineered prompt to avoid irrelevant extra response tokens.
	\item A nested selection mechanism, meaning an agent AI is responsible for making decision to select right model~\cite{plaat2025agentic}.
\end{itemize}

LLMs are also supported by a few techniques that help with reasoning power, enabling task breakdown (impact on caching etc.) and structured output preparation.
Chain-of-Thought is designed to allow sequential reasoning power by analyzing responses received and cached so far to mimic how an intelligent creature decide~\cite{liu2024we}.
This can be impactful when a chain of calls is made to a remote LLM for sub-context members that must contribute to the final output.
Few-shot learning is a technique that extends an input prompt with a few examples to guide a remote model toward a structured response~\cite{xu2024does}.
This structured output preparation can be achieved with other approaches as well.
For instance, injecting some training parameters, resulting in a customized model, or simply using a retry-based attempt with a predefined try-count to eventually receive desired structure. 

There are similar tools that offer code understanding using LLMs.
Authors in~\cite{nam2024using} study the procedure that developers take to send their prompt to LLM.
The IDE-plugin, GILT (Generation-based Information-support with LLM Technology), they have created is capable of injecting contextual information into prompt prior to submission to LLM provider.
Their works generally for all type of questions a developer may want to ask LLM, including implementation tasks.
A few key metrics included when the context added to the prompt such as developer interaction history.
In our work, we suggest a deep, precise context, designed specifically for method comment generation that goes beyond shallow prompt preparation. 

In an another similar effort, authors in~\cite{bappon2024autogenics} recommend enriching prompts for understanding code comment shared in Stack Overflow.
They suggest using context provided in questions to help converging toward better generated comment by LLM.
It is similar to our work as we both intend to provide more precise and relevant context for more accurate responses.

JetBrains AI Assistant is another example of an AI-powered IDE solution, positioning itself as a pioneer in integrating generative AI into modern development environments\footnote{https://www.jetbrains.com/help/idea/ai-assistant-in-jetbrains-ides.html}.
This comprehensive extension provides developers with a wide range of AI features — from code completion and AI chat to intelligent edit suggestions.
SmartDoc, in particular, focuses on delivering an engineered approach for generating comments for given methods.
While we cannot compete with JetBrains AI Assistant as a general-purpose tool that supports various development activities, our work can be seen as an effort to provide an activity-specific solution focused on code comment generation for end users.

\section{SmartDoc}
\label{sec:smartdoc}

SmartDoc, as its name suggests, is a tool that assists developers in comment generation, primarily at the method level\footnote{Available from: \url{https://github.com/vahidetemadi/smartdoc/tree/develop}}.
This tool is delivered as an IntelliJ IDEA plugin and can be considered an AI agent.
This agent is able to produce comments that can equalize with the human-generated text (see Section~\ref{sec:res+eval}, BERTScore).
It leverages a recursive traversal of all callees rooted in the method body to create a temporary memory, enabling taking advantage of RAG principles.
Via following this approach, the eventual request submitted for receiving the comment carries all relevant key contexts combined.
In RAG-powered prompt generation, a contextual data source plays a key role in assisting remote AI to provide accurate response.
Via these settings included in SmartDoc, it can be labeled as a context-aware system.

\subsection{Architecture}
\label{sec:arch}

\begin{figure}
	\centering
	\includegraphics[width=\linewidth]{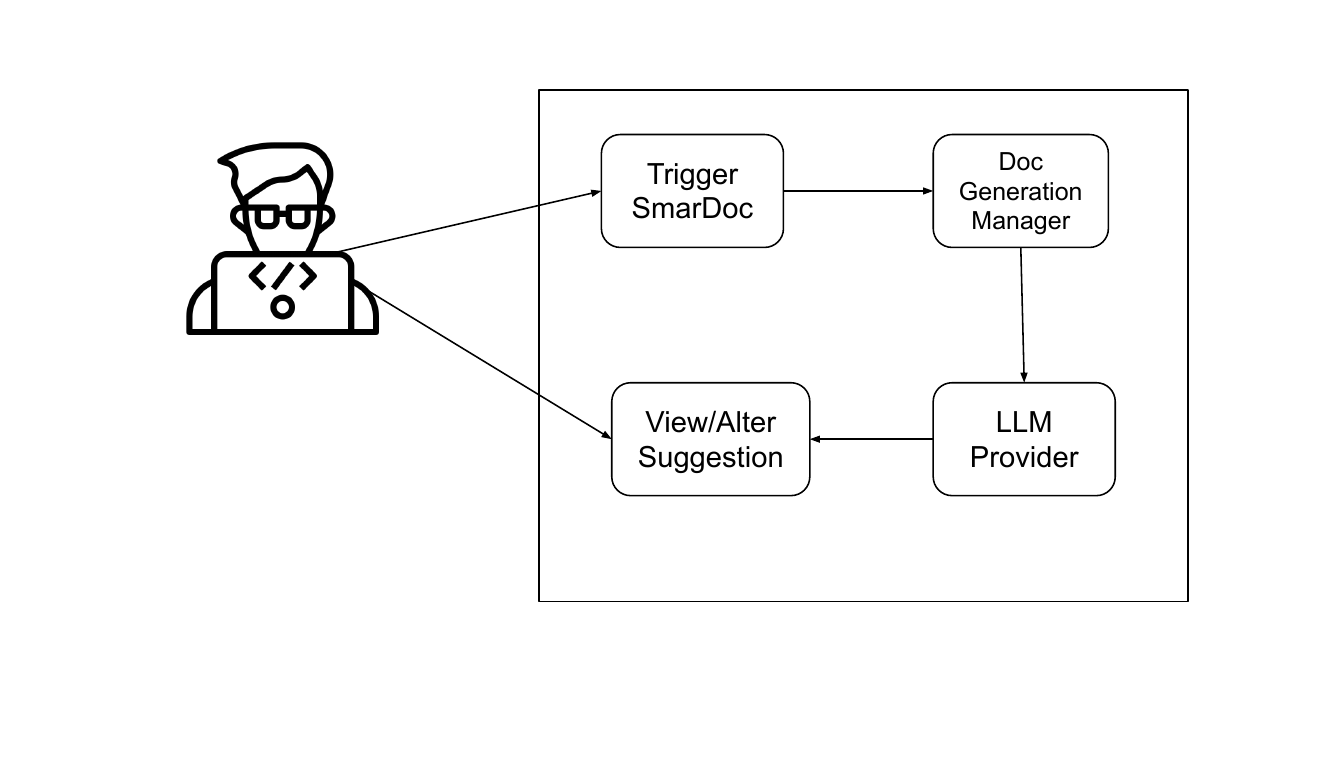}
	\caption{{SmartDoc workflow representing interaction of user with the system and internal data flow}\footnotesize}
	\label{fig:workflow}
\end{figure}

Interaction between SmartDoc and the end-user in the abstract level, as a  workflow, is shown in Figure~\ref{fig:workflow}.
This is a specific flow taken place when a developer interacts with the plugin.
Each request is particularity associated with a method that has a specific signature and input parameters.
Each user's request for a method initiates this workflow that is listened by the doc generator component and ends with a structurally valid JavaDoc style comment (editable by developer as well).
Within this workflow, the doc generator component is responsible for coordinating context provisioning and communication with the remote/local LLM provider.

\begin{figure}
	\centering
	\includegraphics[width=\linewidth]{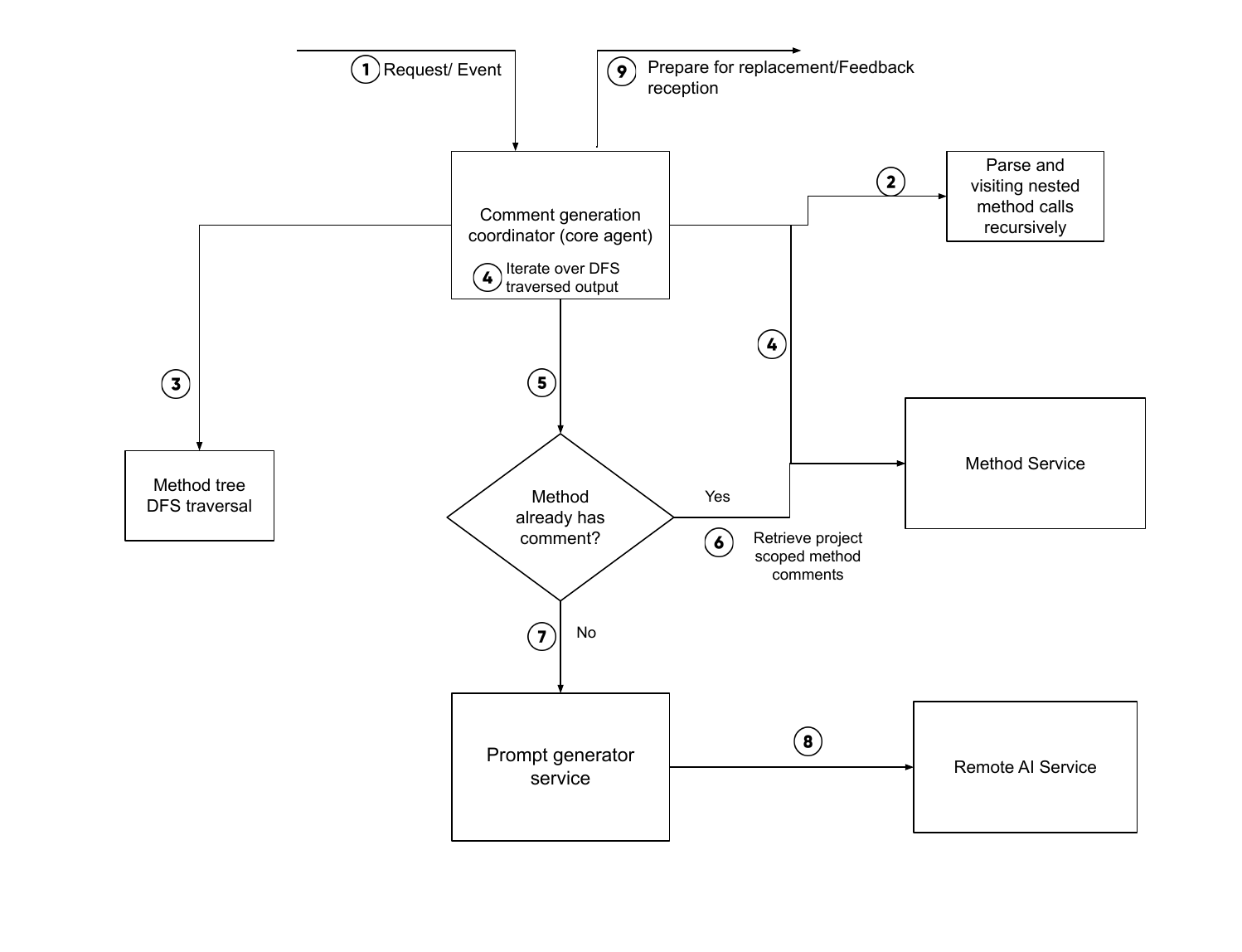}
	\caption{The workflow of the comment generation request}\footnotesize
	\label{fig:sd-thecore}
\end{figure}

At the core, each comment generation request for a given method travels steps 1 to 9 (shown in Figure \ref{fig:sd-thecore}) to generate a comment.
This detailed flow describes the core logic behind processing each request, including all dependencies between services to handle that request.


\begin{figure}
	\centering
	\includegraphics[width=\linewidth]{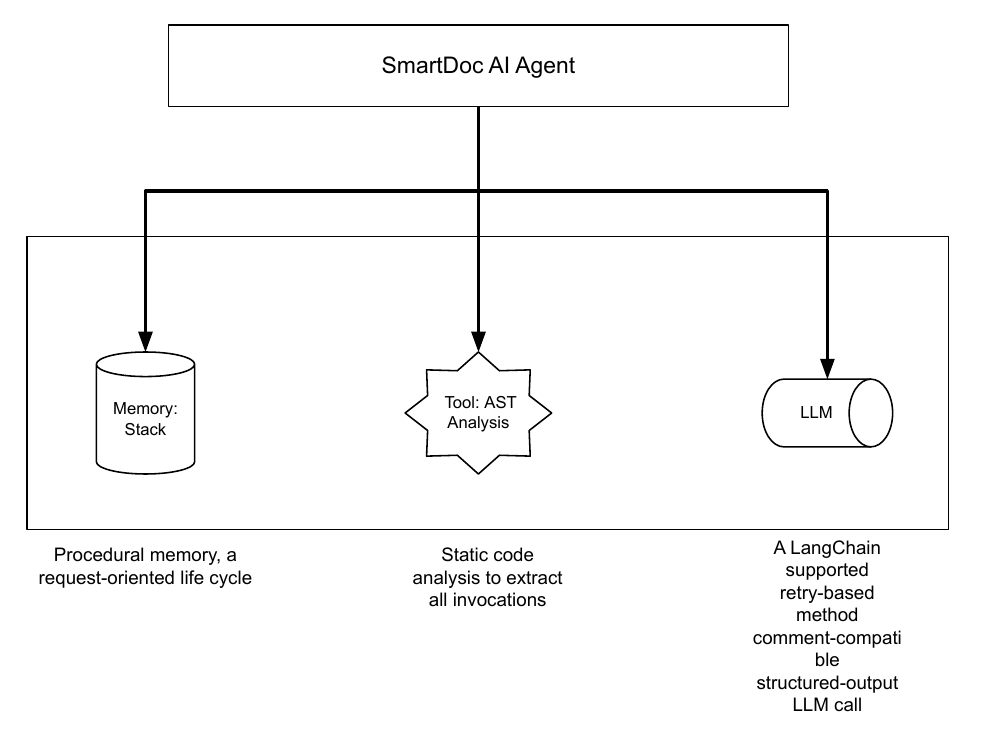}
	\caption{{SmartDoc agent component. It comprises the system core, acting as key runner for producing precise comment.}\footnotesize}
	\label{fig:smartdoc-agent}
\end{figure}

SmartDoc, at its core, is assisted by an AI agent (see Figure~\ref{fig:smartdoc-agent}).
This agent involves a caching memory (suitable for context-preserving and optimized resource usage), that enable a chain-of-thoughts mechanism.
When the agent makes a request to the local/remote LLM\footnote{In this paper, when we call remote or local LLM, it refers to the type of deployment of that model. We anticipated this and use appropriate type of API to make the calls.} and receives a response that contains the JavaDoc-style match, it will consider it for next context inclusion.

\begin{figure}
	\centering
	\includegraphics[width=\linewidth]{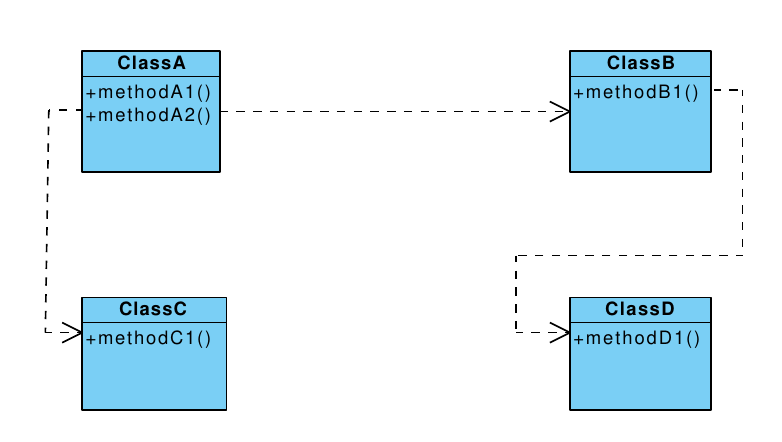}
	\caption{{An example of classes that are coupled via method calls. Assume methodA1 in ClassA has a call to methodA2}\footnotesize}
	\label{fig:cd-example}
\end{figure}

\begin{figure}
	\centering
	\includegraphics[width=\linewidth]{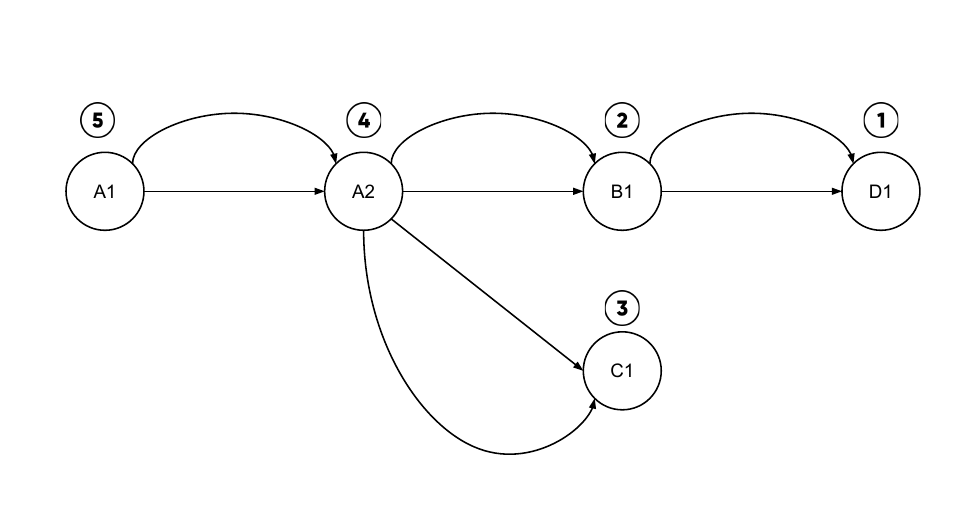}
	\caption{{Call graph and the order of visiting each method, according to the class diagram shown in Figure~\ref{fig:cd-example} }\footnotesize}
	\label{fig:call-graph}
\end{figure}

Figure \ref{fig:cd-example} represents an example of classes that are dependent due to method calls.
These method calls then create a call graph shown in Figure \ref{fig:call-graph}.
We assume it as a tree, meaning circular method calls are not allowed as best practices in software development.
This graph is then traversed DFS to enable context-preparation prior to making the call for parent methods.
In this way, for a given method, when requested to generate a comment for it, all its descendants have an explanation already.
This is a key process that enables offering a RAG-powered solution.  

SmartDoc is working for all the application-level methods definition.
We assume for all public third-party APIs, target LLM understand its explanation, as it has been trained over those publicly available data.
This is a powerful feature of LLMs as they have been trained over millions of data records and input texts, covering a super wide range of publicly available programming languages, SDKs and third-party libraries.
Therefore, no matter how many and in what way third party methods or the programming language core are used, the remote AI can understand and provide interpretation.
At the end, it is only project-internal method calls that are requested to generate new explanation for completing the context.

\subsection{Dealing with structured output}
\label{subsec:structured}

To maximize LLM efficiency in generating responses, it is required to have structured output compatible to the initial requirements.
The final polished output should be a JavaDoc style compatible text, enabling a valid (re)placement for current method comment.
A simple and straightforward solution is instructing through prompts for the required structured output.
LLMs can be provided with an engineered prompt to generate these compatible comments (i.e., instructing via system and user messages in request submission).

During the implementation, we had tow approaches to follow:
(I) Using a retry-based method: defining a max retry count and initiate next request in case of invalid response.
(II) Using a few-shot learning, via injecting specific structure to fine-tune the LLM for only responding valid comments.

For this version, a simple retry-based approach was selected.
It works fine when the model owns remarkable number of parameters and is generalized for a wide range of use cases.
For instance, the DeepSeek remote model responds remarkably accurately before reaching the max-try-count.
In addition, to avoid any inconsistencies, if the response partially contains the desired output, we apply a predefined regex on the response to ensure valid comment replacement.

\section{Results and evaluations}
\label{sec:res+eval}
We developed an application, delivering it as a IntelliJ IDEA plugin, dedicated to Java codebase.
Results in our system means every comment generated for a given method.
This comment text is completely compatible with JavaDoc style standard.
The method comment is applicable when a developer wants to grasp required knowledge to make a change to the project.
Its effectiveness can be recognized when we can evaluate its effect in several ways, ensuring its true impact on the development process.

The current system has a non-deterministic, variable nature, stemming from the AI model which generates comments for the input method.
It means, each time a request is sent a valid and still different response is received, making deterministic test case design difficult. 
This enforces traditional testing useless and needs endeavors for a new test and evaluation design.
All this leads us to focus on different metrics which measures response quality (e.g., relevance, coherence, etc.), as well as deterministic-metrics, such as precision, accuracy and F1.
These metrics can be available using open source implementations that curated for such these purposes.

\subsection{Implementing the evaluation set-up}
To conduct the evaluations, given the current IDE plugin, SmartDoc, we organize the evaluation process in terms of several test cases.
A parameterized test case that is developed to run against different sub modules of the Eclipse Ditto project\footnote{https://eclipse.dev/ditto/}.
We have selected this project since it has a great maintenance and involves implementation with correct, humani integrated method comments.
This test case is in addition to other test cases developed during the development of SmartDoc, as we used various techniques to ensure a maintainable software.

The evaluation of the final output is intended to be done at two levels:
First, to compute how accurate generated comment are in terms of three key metrics.
Second, using a feedback system that receive users' opinion for each generated comment.
Although, this feedback data is expected to be available after first beta release of the plugin.

\subsubsection{Hallucinations (invalid responses)}

Hallucinations refer to invalid or factually incorrect responses generated by an LLM for a given request.
To measure this effect, we can use three commonly adopted community metrics: BERTScore, ROUGE-1, and BLEU~\cite{kumar2025using}.
Although these metrics are not specifically designed to evaluate hallucinations, they provide indicative values that can help assess this phenomenon.
These metrics compute the degree of similarity between the generated output (referred to as the actual) and the human-produced reference output (the expected), based on semantic, exact, and partial token matches.

\begin{figure*}[!ht]
	\centering
	\begin{subfigure}[t]{0.32\linewidth}
		\centering
		\includegraphics[width=\linewidth]{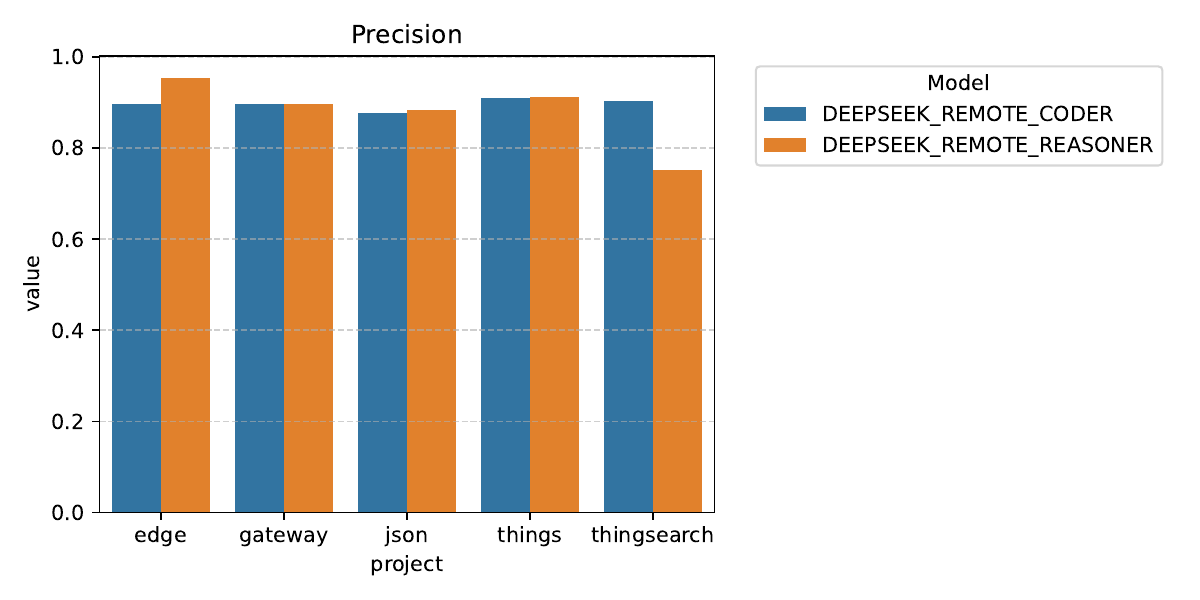}
		\label{fig:bert-f1}
	\end{subfigure}
	\hfill
	\begin{subfigure}[t]{0.32\linewidth}
		\centering
		\includegraphics[width=\linewidth]{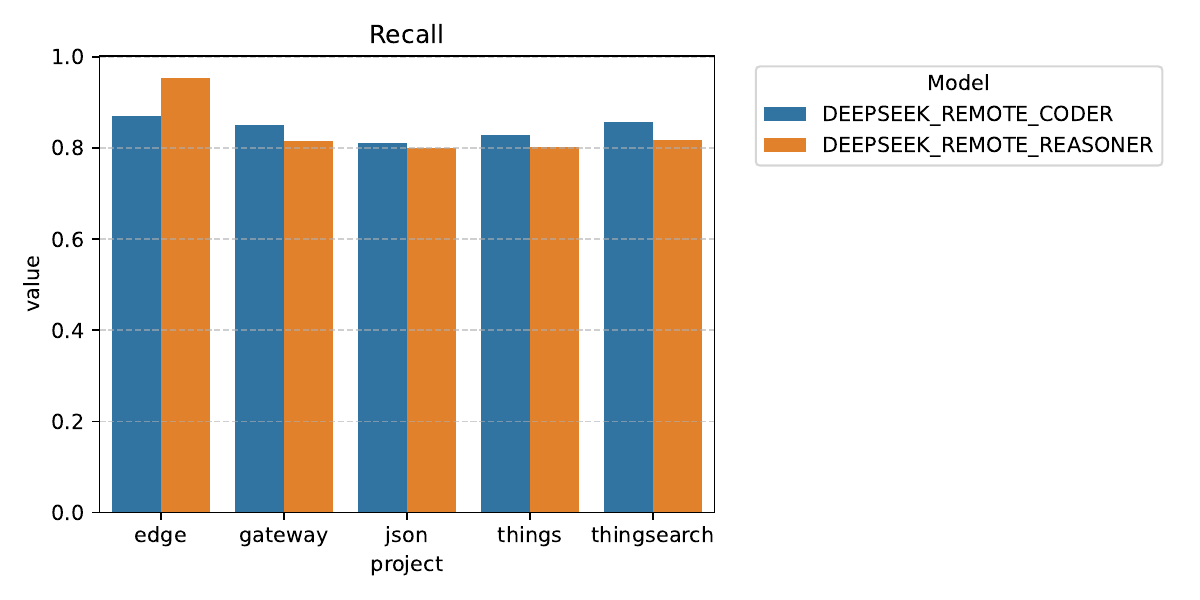}
		\label{fig:bert-precision}
	\end{subfigure}
	\hfill
	\begin{subfigure}[t]{0.32\linewidth}
		\centering
		\includegraphics[width=\linewidth]{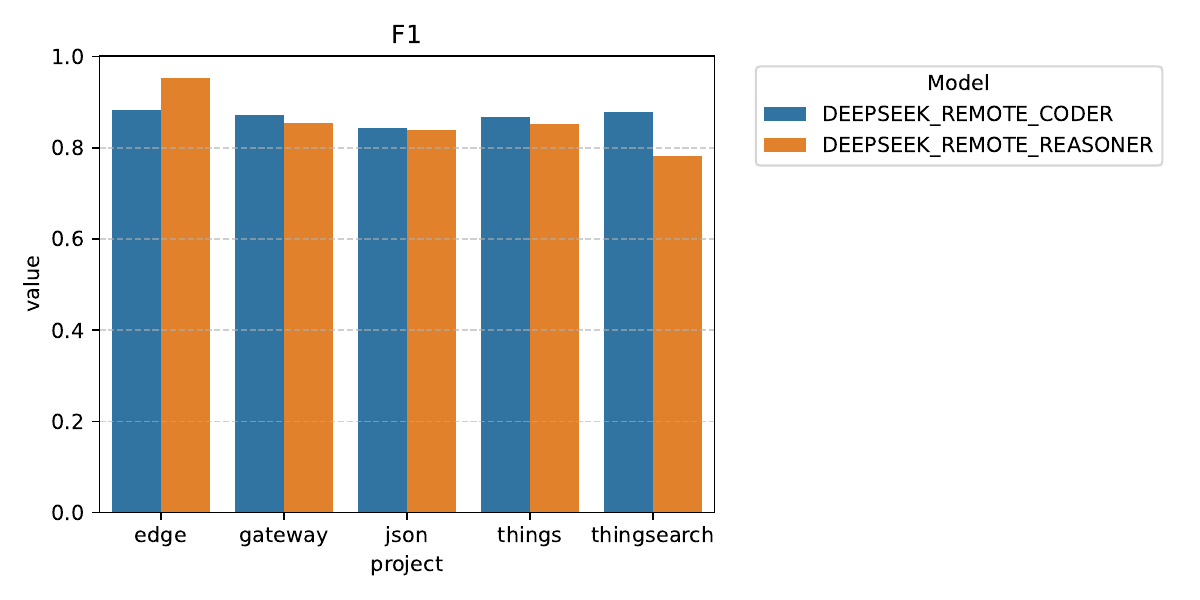}
		\label{fig:bert-recall}
	\end{subfigure}
	\caption{\footnotesize BERTScore metric comparing SmartDoc output with the ground truth}
	\label{fig:bert-all}
\end{figure*}

\emph{BERTScore}:
BERTScore measures semantic similarity between actual and expected outputs, for each given method, using comparison-pair contextual embedding.
Figure~\ref{fig:bert-all} compares the computed metrics for precision, recall and F1 categories for 5 packages of Eclipse Ditto open source project  (applied to minimum of 10 methods per package).

\begin{figure}[H]
	\centering
	\includegraphics[width=\linewidth]{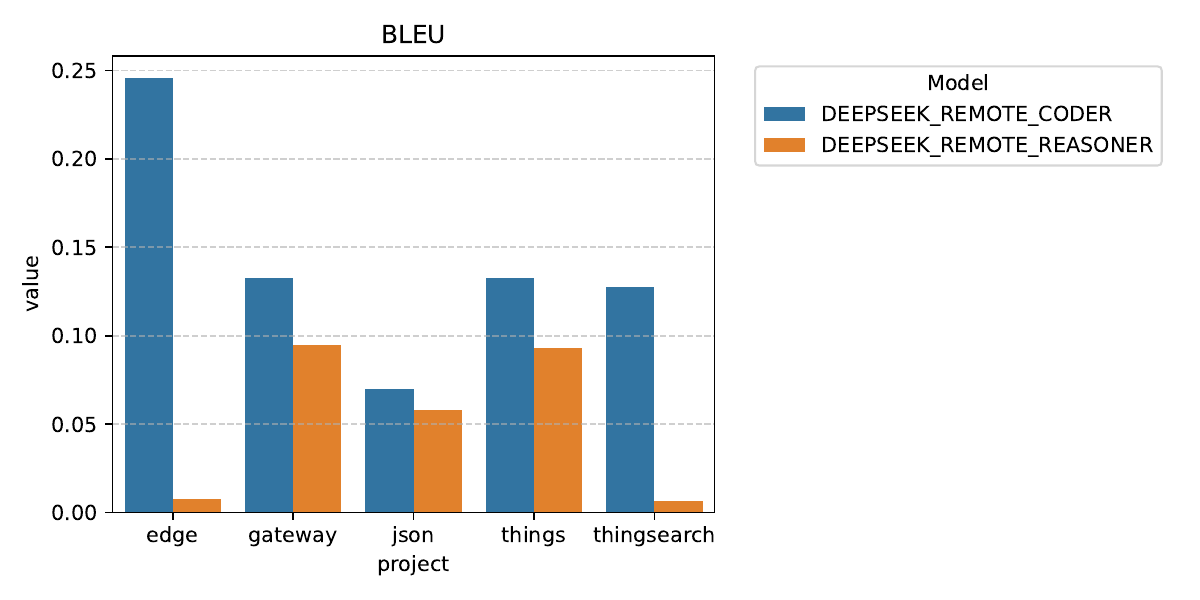}
	\caption{{BLUE metric obtained from comparing SmartDoc with human wrote comments using \it{sacrebleu} package}\footnotesize}
	\label{fig:bleu}
\end{figure}

\emph{BLEU}: 
This metric is often used for machine translation use cases.
However, since our comparison goal has many features in common with those models, we decided to report for this metric as well.
Figure~\ref{fig:bleu} represents the final result for those five packages.

\begin{figure*}[!ht]
	\centering
	\begin{subfigure}[t]{0.32\linewidth}
		\centering
		\includegraphics[width=\linewidth]{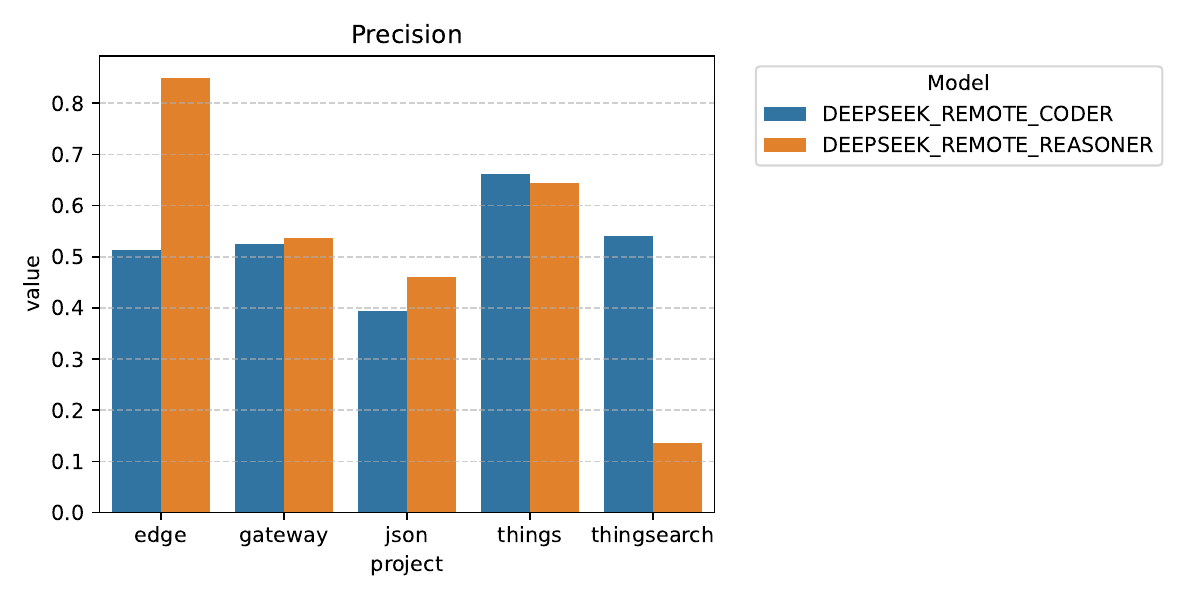}
		\label{fig:rouge-precision}
	\end{subfigure}
	\hfill
	\begin{subfigure}[t]{0.32\linewidth}
		\centering
		\includegraphics[width=\linewidth]{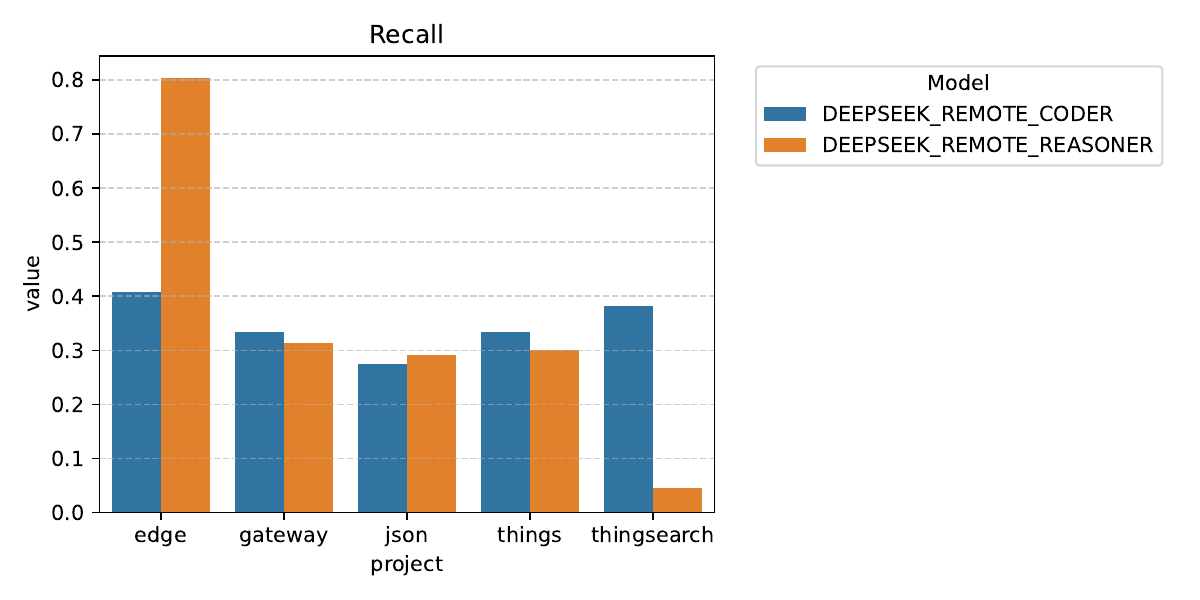}
		\label{fig:rouge-recll}
	\end{subfigure}
	\hfill
	\begin{subfigure}[t]{0.32\linewidth}
		\centering
		\includegraphics[width=\linewidth]{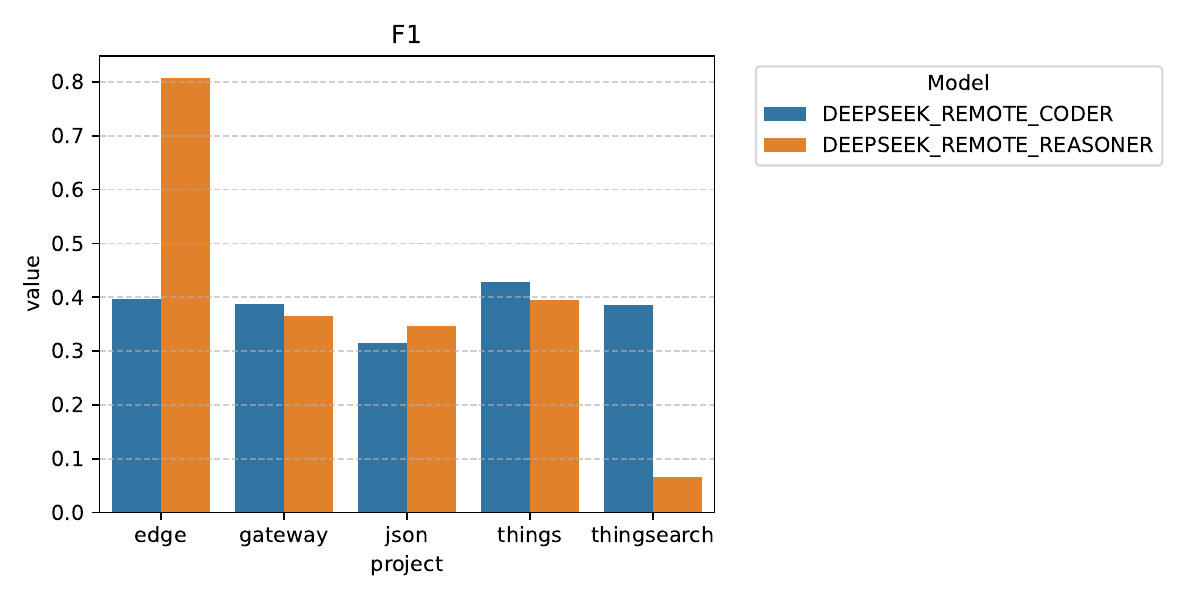}
		\label{fig:rouge-f1}
	\end{subfigure}
	\caption{\footnotesize ROUGE-1 metric comparing SmartDoc output with the ground truth}
	\label{fig:rouge-all}
\end{figure*}

\emph{ROUGE-1 score}:
This metric computes the ratio of uni-grams of actual text (from generated comment) that appear in the expected (from ground truth).
This metric specifically counts number of tokens that is common between actual and expected outputs.
Figure~\ref{fig:rouge-all} lists all comparisons in terms of precision, recall and F1, again for those five packages with minimum 10 methods.

\subsubsection{Feedback collection (satisfied end-user)}

We required to devise an approach to enable user feedback collection.
For this purpose, once a user receives his generated comment (upon the request initiated), the plugin asks for a rate and it is quite optional for the user to send the feedback.
In case of feedback submission, none of user's private information--at any level--is collected, and only the rate and any text provided saved.
This key data can help us to conduct a qualitative study and extend the evaluation.
This analysis actually takes place  after a collection-period done, and can enable us to measure how satisfied the end-users are in general.

\section{Discussion and conclusion}
\label{sec:dis+conc}

We can observe promising results regarding similarity metrics.
BERTScore report is suggesting a range of 0.80 to 0.90 in terms of precision, recall and F1.
ROUGE-1 score, as a partial n-gram comparison, has an average ~0.40 to 0.50 in terms of similar metrics and BLEU, as an exact token based similarity measure offers a value of maximum of 0.25.
This drop in accuracy when moving from BERTScore to BLEU is normal, as it transitions from a semantic similarity measure to exact token matches.

In addition to what evaluated so far, there is a direct relationship between the accuracy of the remote LLM model used and the eventual comments generated.
It means, using a more comprehensive model as the comment generator, we expect better results, and more satisfaction rate.
This is extra to the impact of useful context provided when a call to LLM takes place, meaning a more relevant context can increase the final response relevance.

This plugin is evolvable, meaning there are still new features that can be added while preserving its current reliable functionality.
For instance, support for other languages, plus adding new models to offer the end-users more options.
This plugin enables end-users concurrent comment generation, allowing them to submit multiple requests in a specific time window.
It has minimum interference with text editor events and the users can continue their programming tasks while awaiting responses for comment provision.

We will invest on chain-of-thoughts~\cite{wei2022chain} in more details in the new updates.
In this way, the agent breaks down the engineered prompt preparation given all nested method calls.
In every single step, it can reason and select best explanation for the callees.
It also can be supported by a reasoner to offer its best for context preparation and eventual comment generation.

As mentioned, we enabled a feedback system, allowing users to anonymously submit us their feedback.
This feedback by now allows a five stars rates along with the \textit{LLM model} user selected for his request.
We may decide and offer an agreement asking user to collect the method metadata as well in the newer updates.
It can be effective for analyzing how our method acts in different situation (e.g., number of nested method calls, method length, duration of comment generation, resource utilization)

In this plugin, we focus on static analysis of codebase, only visiting the current snapshot of the codebase.
This can cover methods overloaded and explicit calls.
However,  the method visiting to create call graph becomes more challenging when programmers use specific patterns or design principles that require method overriding or runtime method dispatching.
We plan to cover this feature in the upcoming updates.

\bibliographystyle{ACM-Reference-Format}
\bibliography{library}


\begin{thebibliography}{16}


\ifx \showCODEN    \undefined \def \showCODEN     #1{\unskip}     \fi
\ifx \showISBNx    \undefined \def \showISBNx     #1{\unskip}     \fi
\ifx \showISBNxiii \undefined \def \showISBNxiii  #1{\unskip}     \fi
\ifx \showISSN     \undefined \def \showISSN      #1{\unskip}     \fi
\ifx \showLCCN     \undefined \def \showLCCN      #1{\unskip}     \fi
\ifx \shownote     \undefined \def \shownote      #1{#1}          \fi
\ifx \showarticletitle \undefined \def \showarticletitle #1{#1}   \fi
\ifx \showURL      \undefined \def \showURL       {\relax}        \fi
\providecommand\bibfield[2]{#2}
\providecommand\bibinfo[2]{#2}
\providecommand\natexlab[1]{#1}
\providecommand\showeprint[2][]{arXiv:#2}

\bibitem[Ahmad et~al\mbox{.}(2020)]%
        {ahmad2020transformer}
\bibfield{author}{\bibinfo{person}{Wasi~Uddin Ahmad}, \bibinfo{person}{Saikat
  Chakraborty}, \bibinfo{person}{Baishakhi Ray}, {and} \bibinfo{person}{Kai-Wei
  Chang}.} \bibinfo{year}{2020}\natexlab{}.
\newblock \showarticletitle{A transformer-based approach for source code
  summarization}.
\newblock \bibinfo{journal}{\emph{arXiv preprint arXiv:2005.00653}}
  (\bibinfo{year}{2020}).
\newblock


\bibitem[Bappon et~al\mbox{.}(2024)]%
        {bappon2024autogenics}
\bibfield{author}{\bibinfo{person}{Suborno~Deb Bappon}, \bibinfo{person}{Saikat
  Mondal}, {and} \bibinfo{person}{Banani Roy}.}
  \bibinfo{year}{2024}\natexlab{}.
\newblock \showarticletitle{Autogenics: Automated generation of context-aware
  inline comments for code snippets on programming q\&a sites using llm}. In
  \bibinfo{booktitle}{\emph{2024 IEEE International Conference on Source Code
  Analysis and Manipulation (SCAM)}}. IEEE, \bibinfo{pages}{24--35}.
\newblock


\bibitem[Etemadi et~al\mbox{.}(2022)]%
        {etemadi2022task}
\bibfield{author}{\bibinfo{person}{Vahid Etemadi}, \bibinfo{person}{Omid
  Bushehrian}, {and} \bibinfo{person}{Gregorio Robles}.}
  \bibinfo{year}{2022}\natexlab{}.
\newblock \showarticletitle{Task assignment to counter the effect of developer
  turnover in software maintenance: A knowledge diffusion model}.
\newblock \bibinfo{journal}{\emph{Information and software technology}}
  \bibinfo{volume}{143} (\bibinfo{year}{2022}), \bibinfo{pages}{106786}.
\newblock


\bibitem[Huang et~al\mbox{.}(2025)]%
        {huang2025your}
\bibfield{author}{\bibinfo{person}{Yuan Huang}, \bibinfo{person}{Yinan Chen},
  \bibinfo{person}{Xiangping Chen}, {and} \bibinfo{person}{Xiaocong Zhou}.}
  \bibinfo{year}{2025}\natexlab{}.
\newblock \showarticletitle{Are your comments outdated? Toward automatically
  detecting code-comment consistency}.
\newblock \bibinfo{journal}{\emph{Journal of Software: Evolution and Process}}
  \bibinfo{volume}{37}, \bibinfo{number}{1} (\bibinfo{year}{2025}),
  \bibinfo{pages}{e2718}.
\newblock


\bibitem[Huang et~al\mbox{.}(2023)]%
        {huang2023comparative}
\bibfield{author}{\bibinfo{person}{Yuan Huang}, \bibinfo{person}{Hanyang Guo},
  \bibinfo{person}{Xi Ding}, \bibinfo{person}{Junhuai Shu},
  \bibinfo{person}{Xiangping Chen}, \bibinfo{person}{Xiapu Luo},
  \bibinfo{person}{Zibin Zheng}, {and} \bibinfo{person}{Xiaocong Zhou}.}
  \bibinfo{year}{2023}\natexlab{}.
\newblock \showarticletitle{A comparative study on method comment and inline
  comment}.
\newblock \bibinfo{journal}{\emph{ACM Transactions on Software Engineering and
  Methodology}} \bibinfo{volume}{32}, \bibinfo{number}{5}
  (\bibinfo{year}{2023}), \bibinfo{pages}{1--26}.
\newblock


\bibitem[Kumar et~al\mbox{.}(2025)]%
        {kumar2025using}
\bibfield{author}{\bibinfo{person}{Abhishek Kumar}, \bibinfo{person}{Sandhya
  Sankar}, \bibinfo{person}{Partha~Pratim Das}, {and}
  \bibinfo{person}{Partha~Pratim Chakrabarti}.}
  \bibinfo{year}{2025}\natexlab{}.
\newblock \showarticletitle{Using Large Language Models for multi-level commit
  message generation for large diffs}.
\newblock \bibinfo{journal}{\emph{Information and Software Technology}}
  \bibinfo{volume}{187} (\bibinfo{year}{2025}), \bibinfo{pages}{107831}.
\newblock


\bibitem[Lin et~al\mbox{.}(2024)]%
        {lin2024data}
\bibfield{author}{\bibinfo{person}{Xinyu Lin}, \bibinfo{person}{Wenjie Wang},
  \bibinfo{person}{Yongqi Li}, \bibinfo{person}{Shuo Yang},
  \bibinfo{person}{Fuli Feng}, \bibinfo{person}{Yinwei Wei}, {and}
  \bibinfo{person}{Tat-Seng Chua}.} \bibinfo{year}{2024}\natexlab{}.
\newblock \showarticletitle{Data-efficient Fine-tuning for LLM-based
  Recommendation}. In \bibinfo{booktitle}{\emph{Proceedings of the 47th
  international ACM SIGIR conference on research and development in information
  retrieval}}. \bibinfo{pages}{365--374}.
\newblock


\bibitem[Liu et~al\mbox{.}(2024)]%
        {liu2024we}
\bibfield{author}{\bibinfo{person}{Michael~Xieyang Liu},
  \bibinfo{person}{Frederick Liu}, \bibinfo{person}{Alexander~J Fiannaca},
  \bibinfo{person}{Terry Koo}, \bibinfo{person}{Lucas Dixon},
  \bibinfo{person}{Michael Terry}, {and} \bibinfo{person}{Carrie~J Cai}.}
  \bibinfo{year}{2024}\natexlab{}.
\newblock \showarticletitle{" we need structured output": Towards user-centered
  constraints on large language model output}. In
  \bibinfo{booktitle}{\emph{Extended Abstracts of the CHI Conference on Human
  Factors in Computing Systems}}. \bibinfo{pages}{1--9}.
\newblock


\bibitem[Maalej et~al\mbox{.}(2014)]%
        {maalej2014comprehension}
\bibfield{author}{\bibinfo{person}{Walid Maalej}, \bibinfo{person}{Rebecca
  Tiarks}, \bibinfo{person}{Tobias Roehm}, {and} \bibinfo{person}{Rainer
  Koschke}.} \bibinfo{year}{2014}\natexlab{}.
\newblock \showarticletitle{On the comprehension of program comprehension}.
\newblock \bibinfo{journal}{\emph{ACM Transactions on Software Engineering and
  Methodology (TOSEM)}} \bibinfo{volume}{23}, \bibinfo{number}{4}
  (\bibinfo{year}{2014}), \bibinfo{pages}{1--37}.
\newblock


\bibitem[Nam et~al\mbox{.}(2024)]%
        {nam2024using}
\bibfield{author}{\bibinfo{person}{Daye Nam}, \bibinfo{person}{Andrew Macvean},
  \bibinfo{person}{Vincent Hellendoorn}, \bibinfo{person}{Bogdan Vasilescu},
  {and} \bibinfo{person}{Brad Myers}.} \bibinfo{year}{2024}\natexlab{}.
\newblock \showarticletitle{Using an llm to help with code understanding}. In
  \bibinfo{booktitle}{\emph{Proceedings of the IEEE/ACM 46th International
  Conference on Software Engineering}}. \bibinfo{pages}{1--13}.
\newblock


\bibitem[Niazi et~al\mbox{.}(2023)]%
        {niazi2023investigating}
\bibfield{author}{\bibinfo{person}{Tahira Niazi}, \bibinfo{person}{Teerath
  Das}, \bibinfo{person}{Ghufran Ahmed}, \bibinfo{person}{Syed~Muhammad Waqas},
  \bibinfo{person}{Sumra Khan}, \bibinfo{person}{Suleman Khan},
  \bibinfo{person}{Ahmed~Abdelaziz Abdelatif}, {and} \bibinfo{person}{Shaukat
  Wasi}.} \bibinfo{year}{2023}\natexlab{}.
\newblock \showarticletitle{Investigating novice developers’ code commenting
  trends using machine learning techniques}.
\newblock \bibinfo{journal}{\emph{Algorithms}} \bibinfo{volume}{16},
  \bibinfo{number}{1} (\bibinfo{year}{2023}), \bibinfo{pages}{53}.
\newblock


\bibitem[Plaat et~al\mbox{.}(2025)]%
        {plaat2025agentic}
\bibfield{author}{\bibinfo{person}{Aske Plaat}, \bibinfo{person}{Max van
  Duijn}, \bibinfo{person}{Niki van Stein}, \bibinfo{person}{Mike Preuss},
  \bibinfo{person}{Peter van~der Putten}, {and} \bibinfo{person}{Kees~Joost
  Batenburg}.} \bibinfo{year}{2025}\natexlab{}.
\newblock \showarticletitle{Agentic large language models, a survey}.
\newblock \bibinfo{journal}{\emph{arXiv preprint arXiv:2503.23037}}
  (\bibinfo{year}{2025}).
\newblock


\bibitem[Sauvola et~al\mbox{.}(2024)]%
        {sauvola2024future}
\bibfield{author}{\bibinfo{person}{Jaakko Sauvola}, \bibinfo{person}{Sasu
  Tarkoma}, \bibinfo{person}{Mika Klemettinen}, \bibinfo{person}{Jukka Riekki},
  {and} \bibinfo{person}{David Doermann}.} \bibinfo{year}{2024}\natexlab{}.
\newblock \showarticletitle{Future of software development with generative AI}.
\newblock \bibinfo{journal}{\emph{Automated Software Engineering}}
  \bibinfo{volume}{31}, \bibinfo{number}{1} (\bibinfo{year}{2024}),
  \bibinfo{pages}{26}.
\newblock


\bibitem[Sun et~al\mbox{.}(2024)]%
        {sun2024source}
\bibfield{author}{\bibinfo{person}{Weisong Sun}, \bibinfo{person}{Yun Miao},
  \bibinfo{person}{Yuekang Li}, \bibinfo{person}{Hongyu Zhang},
  \bibinfo{person}{Chunrong Fang}, \bibinfo{person}{Yi Liu},
  \bibinfo{person}{Gelei Deng}, \bibinfo{person}{Yang Liu}, {and}
  \bibinfo{person}{Zhenyu Chen}.} \bibinfo{year}{2024}\natexlab{}.
\newblock \showarticletitle{Source code summarization in the era of large
  language models}.
\newblock \bibinfo{journal}{\emph{arXiv preprint arXiv:2407.07959}}
  (\bibinfo{year}{2024}).
\newblock


\bibitem[Wei et~al\mbox{.}(2022)]%
        {wei2022chain}
\bibfield{author}{\bibinfo{person}{Jason Wei}, \bibinfo{person}{Xuezhi Wang},
  \bibinfo{person}{Dale Schuurmans}, \bibinfo{person}{Maarten Bosma},
  \bibinfo{person}{Fei Xia}, \bibinfo{person}{Ed Chi}, \bibinfo{person}{Quoc~V
  Le}, \bibinfo{person}{Denny Zhou}, {et~al\mbox{.}}}
  \bibinfo{year}{2022}\natexlab{}.
\newblock \showarticletitle{Chain-of-thought prompting elicits reasoning in
  large language models}.
\newblock \bibinfo{journal}{\emph{Advances in neural information processing
  systems}}  \bibinfo{volume}{35} (\bibinfo{year}{2022}),
  \bibinfo{pages}{24824--24837}.
\newblock


\bibitem[Xu et~al\mbox{.}(2024)]%
        {xu2024does}
\bibfield{author}{\bibinfo{person}{Derek Xu}, \bibinfo{person}{Tong Xie},
  \bibinfo{person}{Botao Xia}, \bibinfo{person}{Haoyu Li},
  \bibinfo{person}{Yunsheng Bai}, \bibinfo{person}{Yizhou Sun}, {and}
  \bibinfo{person}{Wei Wang}.} \bibinfo{year}{2024}\natexlab{}.
\newblock \showarticletitle{Does Few-Shot Learning Help LLM Performance in Code
  Synthesis?}
\newblock \bibinfo{journal}{\emph{arXiv preprint arXiv:2412.02906}}
  (\bibinfo{year}{2024}).
\newblock


\end{thebibliography}
\end{document}